# Unveiling the structural, chemical state, and optical band-gap evolution of Ta-doped epitaxial SrTiO$_3$ thin films using first-principles calculations and spectroscopic ellipsometry

Shammi Kumar[1], Raja Sen[2*], Mamta Arya[1], Sankar Dhar[1], and Priya Johari[1*]

1. Department of Physics, School of Natural Science, Shiv Nadar Institute of Eminence (Deemed to be University), Gautam Buddha Nagar, Uttar Pradesh 201314, India.
2. Sorbonne Université, Museum National d'Histoire Naturelle, UMR CNRS 7590, Institut de Minéralogie, de Physique des Matériaux et de Cosmochimie, 4 place Jussieu, F – 75005 Paris, France

*Corresponding Authors: rs190@snu.edu.in, priya.johari@snu.edu.in

**ABSTRACT**

In this report, the optical properties of Ta doped SrTiO$_3$ (STO) due to its potential in transparent conducting oxides (TCOs) is explored by a combination of theoretical studies based on density functional theory and spectroscopic ellipsometry. To achieve this theoretically, we vary the concentration of Ta from 0 - 12.5% in SrTi$_{1-x}$Ta$_x$O$_3$ system by substitutional doping and report its effect on the resulting structural, chemical, electronic, chemical, and optical properties. Additionally, we perform band unfolding to shed light on the true nature of optical transitions due to Ta doping. We verify these results experimentally by fabricating epitaxial SrTi$_{1-x}$Ta$_x$O$_3$ thin films ( x = 0 - 5%) by pulsed laser deposition and obtain the optical dielectric properties of the system with the help of spectroscopic ellipsometry. By combining theoretical and experimental studies, we provide evidence that the band gap of STO increases due to Ta doping while also enhancing its electronic properties. The findings of our study offer an extensive understanding of the intricacies associated with elemental doping in perovskite oxides and propose strategies for addressing obstacles associated with TCOs.

## I. INTRODUCTION

A wide variety of conventional devices, including light-emitting diodes (LEDs), solar cells, touch screens, and smart glasses, rely on the use of transparent-conducting electrode that combines high conductivity with excellent transparency in the visible spectrum [1–6]. In the quest for high-performance transparent-conducting materials (TCMs), various oxides, such as tin-doped indium oxides (ITO) [7], Al-doped ZnO [8], Sb-doped $SnO_2$ [9], amid others, have been the subject of extensive research over the past two decades. Out of these materials, ITO stands out as an ideal candidate for widespread application in the optoelectronic industry due to its tunable band gap of 3.27 to 3.75 eV, minimal absorption peaks in the UV-Visible-NIR regions, and fabrication compatibility with present semiconductor technologies. However, multiple reports have emphasized the urgent need to explore alternatives to ITO due to the supply risk of indium [10,11]. Among the plethora of materials explored as alternatives to ITO, $SrTiO_3$ (STO) has shown great promise as a TCM because of its compatible physical properties, ease of fabrication, cost-effectiveness, and abundant availability in nature [12]. Pristine STO exhibits very high transmittance in the visible region of the spectrum (~ 90 %), accompanied by a large band gap of 3.25 eV [13,14]. However, owing to its inherent electrically insulating nature, it is essential to introduce doping in STO to improve its electrical conductivity, effectively converting it from an insulator to a metal oxide. Beyond its possible application in the optoelectronic industry, doped STO also holds substantial potential in various other fields, including photocatalysis, thermoelectricity, hydrogen fuel cells, gas sensing, and low-temperature superconductivity [15–20].

In the pursuit of exploring promising dopants to create n-type STO, several works have explored the potential of Ta, whether as a primary or a co-dopant at the Ti site of STO. For instance, by conducting theoretical studies on the 12.5% Ta-doped STO system, Hou *et al.* [21] found that Ta effectively donates its electrons to the conduction band of STO without modifying its electronic band gap. Their study further revealed that akin to the pristine system, Ta-doped STO is also non-magnetic. On the contrary, Modak *et al.* and Liu *et al.* reported that the (Ta, Rh) and (Ta, N) co-dopant pair reduces the band gap of STO [22,23], consequently enhancing its visible light activity. Moriga *et al*. [24] have shown that Ta-doped cubic STO is stable only in a broad concentration range of Ta doping (~20%), which has been confirmed independently by Yaremchemko *et al.* [25]. Nevertheless, it is crucial to mention that most of these previous studies have primarily focused on understanding the electronic, optical, and transport properties of Ta-doped $SrTiO_3$ for photocatalysis and thermoelectric device applications. To effectively utilize STO as a TCM, it is crucial to comprehend the complexities associated with the variations in band gap, transmittance, and electrical conductivity of STO in relation to different concentrations of Ta doping. In a connected previously reported work [26,27], we have shown that 5% Ta-doped STO thin film possesses a remarkable transmittance (~ 80-90%), a low electrical resistivity in the range of 5 mΩ-cm, and a comparably high charge-carrier mobility of 5 $cm^2V^{-1}s^{-1}$ at room temperature, making it an exciting material for transparent conducting applications. However, before integrating Ta-doped STO into optoelectronic-based devices, it is crucial to thoroughly investigate the impact of Ta-doping on the variation of the band gap of STO. This aspect of research has not been extensively studied and calls for detailed exploration through both theory and experiment.

Furthermore, it is worth noting that previous theoretical reports on electron-doped STO have not been able to adequately explain the observed trends in experimentally obtained band gaps. For instance, Chen *et al.* [28] have demonstrated an indirect nature upon electron doping of STO, whereas Hou *et al.* [21] have reported a direct transition. This issue arises from the use of supercell approach to simulate doping behaviour, which folds the band structure in the first Brioullin zone. Consequently, the results obtained might not provide an accurate description of the band structure which can be seen experimentally from ARPES studies [29]. This dispute regarding the characteristic of band dispersion can be resolved through employing the methodology proposed by Popescu *et al.* and others [30–35] to *unfold* the band structure, revealing correct band dispersion and determination of the nature of band gap. Another key consideration during doping is the interplay between various factors such as the Moss-Burstein effect, band gap renormalization, electron-impurity interaction, and electron-phonon interaction [36–38]. Due to band filling, the MB effect widens the band gap in doped systems [39,40]; however, other factors counteract this MB effect and cause the band gap to contract [15,41]. In order to ascertain the complications introduced by doping and validate the theoretical models, experimental determination of the optical properties is also crucial.

Hence in this work, we combine theoretical and experimental studies to understand the effect of Ta doping on the optical properties of STO for fabrication of new generation of TCOs. The theoretical studies are first reported in detail to shed light on the effect of Ta on the structural, electronic, chemical, and finally the optical properties. On the experimental side, thin films of epitaxial Ta doped STO are grown by pulsed laser deposition and consequently optical properties are obtained by spectroscopic ellipsometry. The evolution of band gaps obtained theoretically is compared with the experimental results. The comparison of theoretical and experimental results illustrates the band gap variation due to doping in STO saturates at ~10% doping and larger doping has no significant effect on the transparency of STO in the visible spectrum due to the inter-competing effects as mentioned in the preceding paragraph .

## II.  METHODS

### A. Computational Methods

We carry out the first principles based density functional theory (DFT) calculations using the Vienna ab initio simulation package (VASP) based on projector augmented wave (PAW) [42] pseudopotentials. We chose a cut-off kinetic energy of 600 eV, which is 1.5 times the energy cutoff of O ions. Both the lattice and bond length are allowed to relax till the energy convergence criterion of $10^{-6}$ eV with Hellman-Feynman forces on each atom at $10^{-3}$ eV/Å is attained. Gamma centered Monkhorst-Pack grid with a resolution of 0.02 Å$^{-1}$ is chosen for obtaining a higher accuracy in the total energy. Four different structures corresponding to different level of doping are studied. A primitive of STO with space group $Pm\bar{3}m$ contains 5 atoms and has a lattice parameter of 3.905 Å (Figure 1(a)). The k-point mesh for this is 13×13×13 for maintaining a 0.02 Å$^{-1}$ resolution. To simulate the doped system, one Ti atom is replaced by one Ta atom in the supercells of size 4×4×4 (320 atoms), 3×3×3 (135 atoms), and 2×2×2 (40 atoms) to get a corresponding doping percentage of 1.6%, 3.7%, 12.5%, respectively. The k-point mesh is chosen accordingly to maintain this resolution, i.e., 4×4×4, 5×5×5, and 7×7×7 for 1.6%, 3.7%, and 12.5%, respectively. Two different exchange correlation functionals, namely PBE [43] and HSE06 [44,45] are utilized to obtain the electronic properties. The *band unfolding* is done with the use of BandUP [46,47] package developed by Medeiros *et al*. The frequency-dependent optical studies are performed to obtain the absorption spectrum.

### B. Experimental Methods

Epitaxial films of 0, 2%, and 5% Ta doped STO are grown on $LaAlO_3$ (001) substrates with the help of pulsed laser deposition (PLD). All these samples have a same growth condition i.e., a growth temperature ($T_g$) of 750°C, laser energy density of 1.5 $J/cm^2$ and oxygen partial pressure ($pO_2$) of $10^{-5}$ mbar. The thicknesses of these films range are in ~100-130 nm range. Measurements of optical properties are carried out with the help of variable angle spectroscopic ellipsometer (VASE) (model M-2000-UI, J. A. Woollam, USA) in the standard geometry. The schematic of VASE is given in Figure 1(b). The measurement is taken at 3 different angles of 55°, 60° and 65° to get reliant optical characteristics. The modelling is done by the CompleteEASE software package. Other details and results on the structural, chemical and electronic properties has been reported elsewhere [26,27].

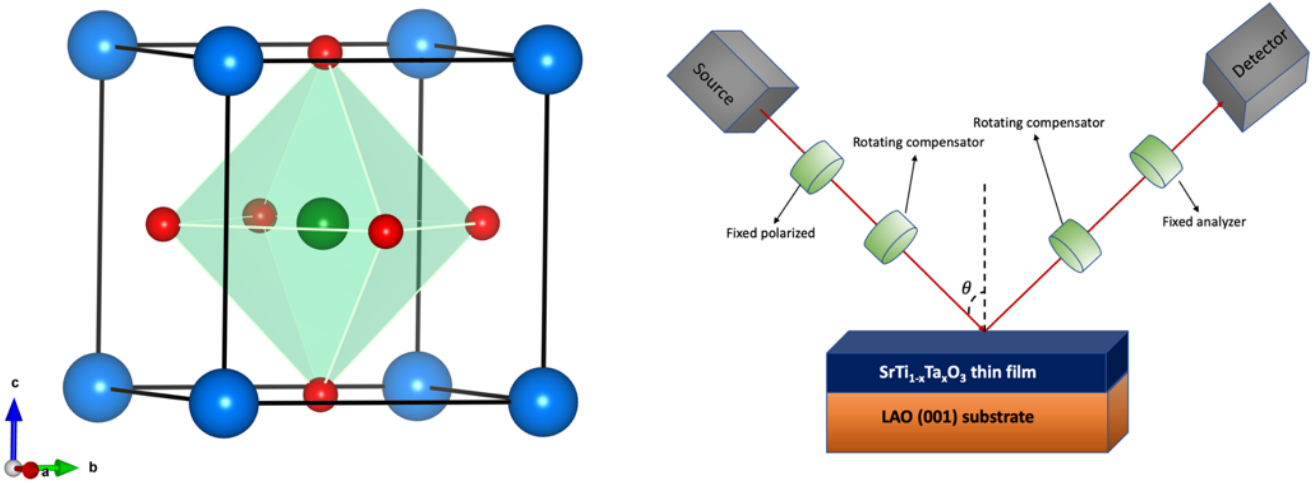

**Figure 1**: (a) Unit cell of STO, and (b) schematic of spectroscopic ellipsometer measurement.

## III. RESULTS AND DISCUSSION

### A. Structural and chemical properties of pure and Ta doped STO

To gain insights into the impact of Ta doping on the structural stability of STO, we begin by computing dopant formation energy for Ta substitution in STO at different concentrations of doping, ranging from 1.6% to 12.5 at.%. The primitive cell structure of STO is shown in Fig. 1(a). At room temperature, the compound crystallizes in a cubic perovskite structure with space group $Pm\bar{3}m$. It can be seen from Fig. 1(a) that the cubic structure of STO exhibits pure centrosymmetry with the Ti atom residing at the center of the octahedron formed by six $O^{2-}$ anions, while the Sr atoms occupy the eightfold corner-sharing positions. Within the DFT-PBE theory, we obtain the lattice constant of STO to be 3.942 Å which shows a very good agreement with experiments (3.905 Å), as well as previous reports [48]. This primitive structure of STO is thus adopted as the basis for designing all the Ta doped supercells in the present study. Considering the fact that Ta atoms preferentially occupy Ti sites rather than Sr sites [23], this study exclusively concentrates on the substitution of Ti atoms with Ta. Herein, we particularly choose three Ta-doped systems of STO with dopant concentrations of 1.6, 3.7, and 12.5 at.%. These systems are realized theoretically by employing three different supercells of STO, i.e., 2×2×2, 3×3×3, and 4×4×4 supercells, and replacing one Ti atom with Ta in each supercell. Prior to the analysis of the stability and structural properties, the doped structures are unbiasedly relaxed until the forces between each pair of atoms are less than 1 meV/Å. The dopant formation energy ($E_f$) is then calculated using the following equation [28]

$$E_f = (E_{doped} + \mu^{Ti}) - (E_{pure} + \mu^{Ta}) \qquad (1)$$

where $E_{doped}$ and $E_{pure}$ represent the total energy of the supercell with and without the Ta dopant, receptively, and $\mu_{Ti}$ ($\mu_{Ta}$) is the chemical potential of Ti (Ta) atom, defined as the energy per atom of its room-temperature stable phase having symmetry $P6_3/mmc$ ($P4_2/mnm$) [49]. The corresponding results together with the calculated lattice parameter and nearest neighbor atomic bond length for both pure and examined doped systems are given in Table 1. Theoretically, a lower dopant formation energy signifies a higher stability of the doped structure and an increased solubility of the dopant within the lattice. In light of this, the small positive dopant formation energies of the 1.6% and 3.7% Ta-doped STO systems, in comparison to the 12.5% Ta-doped STO, suggest that although not occurring spontaneously, these dopant concentrations can be easily introduced into the pristine system using standard experimental techniques. Although, it is important to mention here that the dopant formation energy of the 12.5% Ta-doped STO is much smaller than the 12.5% M-doped STO (where M = Mn, Ru, Rh, Pd, Ir, and Pt [15,16,21,28,50–54]); among them, some were experimentally studied [55–57]. Further, Ta doping as high as ~20% has been experimentally fabricated and studied in ref. [24,25].

The increase in doping concentrations leads to an increase in the lattice parameter of the pure STO. Indeed, our study finds that in accordance with Vegard's law, the lattice parameters exhibit a linear relationship with the carrier concentrations, as illustrated in Fig. S2 of the Supporting Information (SI).

Table 1: The obtained lattice parameter, formation energy, and the bond lengths of Ta-doped STO system at different concentration. The effect of doping on the local structure distortion is seen clearly due to the introduction of Ta.

| System (% doped Ta) | Lattice Parameter (Å) | Formation Energy (Kcal/mol) | Sr-O Bond length (Å) | | | Ti-O Bond Length (Å) | | | Ta-O (Å) |
|---|---|---|---|---|---|---|---|---|---|
| | | | 1st Neighbour | 2nd Neighbour | 3rd Neighbour | 1st Neighbour | 2nd Neighbour | 3rd Neighbour | |
| Pure | 3.942 | -- | 2.787 | -- | -- | 1.971 | -- | -- | -- |
| 1.6% | 3.944 | 6.262 | 2.837 | 2.788 | 2.775 | 1.930 - 2.005 | 1.970-1.976 | 1.969-1.973 | 1.981 |
| 3.7 % | 3.948 | 7.581 | 2.845 | 2.799 | 2.787 | 1.942-1.997 | 1.972-1.978 | 1.969-1.972 | 1.981 |
| 12.5 % | 3.960 | 10.934 | 2.850 | 2.801 | 2.801 | 1.983 | 1.978 | 1.978 | 1.982 |

In our previous work, we demonstrated that the theoretical increase of the lattice parameter of Ta doped STO (1.6% to 12.5% doped) is consistent with experimental measurements and is the combined result of three competing effects: (i) the size of the Ta dopant, (ii) the deformation potential caused by hydrostatic strains, and (iii) the density of free and compensating carriers [27]. Although, one can find from Table 1 that the lattice parameter of doped systems exhibits only a little increase compared to the pristine STO, indicating no significant structural changes in STO upon low Ta doping. Our study, as expected, finds that due to the similarity in ionic radii between Ti and Ta (60.5 pm and 64.0 pm [58], respectively), the Ta-O bond length (1.981 Å) in the doped system is very similar to that of Ti-O (1.971 Å), which consequently leads to no significant change in the lattice parameter of the pure system following the doping process. However, similar to previous theoretical works [23,28], our DFT calculations identify a noticeable lattice distortion near the dopant site due to the alternation of Ti-O bond length and Ti-O-Ti bond angle. Under the doping of Ta, the nearest neighbour Ti-O bond length and the corresponding TiO$_6$ octahedra undergo shrinkage to accommodate the slightly larger Ta-ion in the Ti site, leading to stretching and shrinking of

Ti-O bonds away and towards the defect, respectively (see Fig. S2 in SI). Although, one can further see from Table 1, this bond length alternation is very localized and diminishes rapidly away from the defect center. The same is true for Sr-O bond length which increases around the dopant site due to the shrinkage of the Ti-O bond length.

**Table 2:** Bader charge analysis of the Ti and Ta ions in different supercells showing the change in oxidation states of Ti as result of Ta doping.

| System (Ta (%)) | Ti ($\bar{e}$) | Ta ($\bar{e}$) |
|---|---|---|
| Pure (No Ta) | 3.515 | -- |
| 1.6 % | 3.45 | 4.85 |
| 3.7% | 3.39 | 4.72 |
| 12.5% | 3.25 | 4.43 |

We next turn our attention to explore the impact of Ta doping on the change of chemical state of the ions in STO. In our previous experimental study of the electronic properties of the Ta doped STO (2% to 10% doped), we showed that the measured room-temperature charge carrier density of the examined systems is about 46-75% of the nominal charge carrier density, where the latter is calculated under the assumption that each Ta ion present in the system contributes one free electron to the system [27]. These charge carriers compensation originates from different types of anion and cation defects [25,59], as confirmed by previous XPS studies [26]. We reported that under doping of Ta and in the presence of O site defects, some Ti changes its valence state from $Ti^{4+}$ to $Ti^{3+}$, and even to $Ti^{2+}$. Besides, Ta is also found to exist in a combination of two oxidation states, namely $Ta^{5+}$ and $Ta^{4+}$. $Ta^{4+}$ contributes one free electron to the system, whereas the electrons around $Ta^{5+}$ are highly localized, as a result of which the samples are not completely ionized.

To further establish the effect of doping on the change of oxidation state of the Ti and Ta ions in doped STO, we here perform the Bader charge analysis of the atoms for both pure and doped systems, as reported in Table 2. Our study finds that the increase of Ta doping in STO slightly reduces the average charge on Ti ions as compared to the pristine system, indicating that some Ti atoms will move from +4 to some lower oxidation states, such as $Ti^{3+}$, due to the doping of Ta. The reduction of oxidation state of Ti compensates any free electrons present in the system and therefore decreases the conductivity of the system. The Bader charge of Ta is also found to decrease monotonically as the concentration of Ta increases in the system, with values changing from 4.85 $\bar{e}$ at 1.6% doping to 4.43 $\bar{e}$ at 12.5% doping. This reduction indicates that with the increase of doping, some Ta atoms will change its valency from 5+ to a lower oxidation state, such as $Ta^{4+}$, and release some electrons to the system. Given that the relative decrease in the Bader charge of Ta with increasing doping is comparatively greater than that of Ti, the free carrier density rises as doping increases. This, in turn, establishes why the measured charge compensation in Ta doped STO system decreases with the increase of doping.

### B. Electronic Properties of pure and Ta doped STO

To unveil the impact of Ta doping on the electronic properties of STO, we next conduct a comparative first-principles study of the band structures and density of states of pristine and doped STO. It is well known that the accuracy of DFT in studying the electronic properties of matter is largely dependent on the choice of the exchange-correlation (xc) functional. In this study, we therefore considered two xc-functionals, namely PBE and HSE06, to assess their capability in accurately predicting the optoelectronic properties of STO, as compared to experimental measurements. Figure 2 represents the electronic band structures of pure STO obtained using both two xc-functionals. We find that the nature of band dispersion of STO throughout the whole Brillouin zone (Fig. 2(a) and (b)) remains consistent for both the functionals and matches well with the experimental band structure of STO studied by ARPES [29]. However, as expected, the PBE-functional significantly underestimates the electronic band gap of STO (1.87 eV,

R→Γ), while HSE06 predicts the band gap to be 3.16 eV, closely matching the experimental band gap of 3.25 eV. Therefore, unless explicitly mentioned, we consider the HSE06 functional for studying the electronic properties of the pure and doped STO.

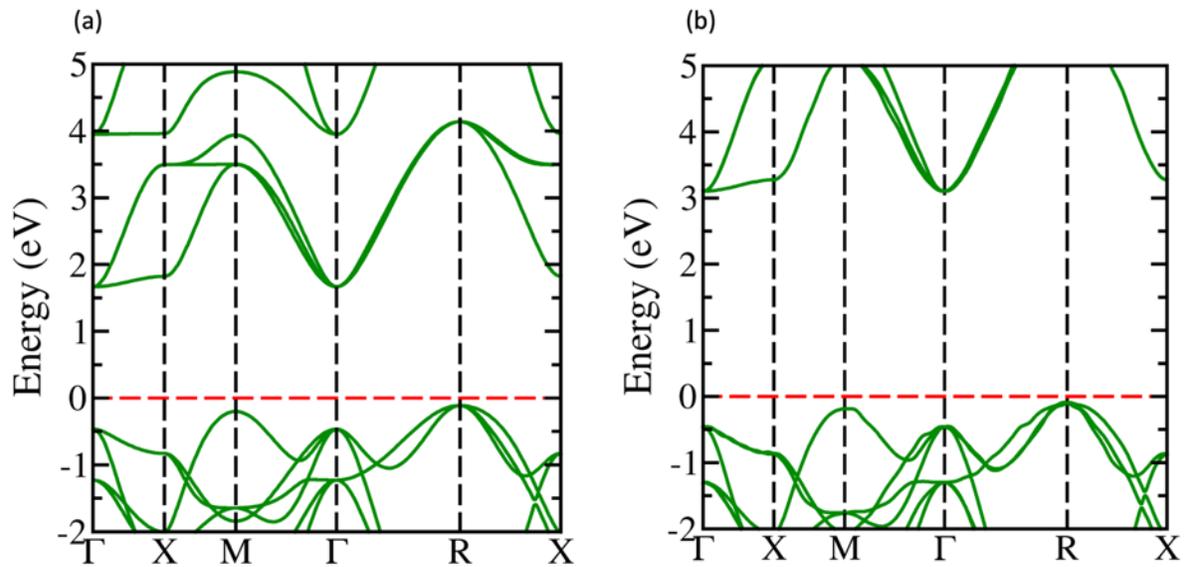

**Figure 2**: Band structure of pure SrTiO$_3$ as obtained from a) PBE and b) HSE06 functional. The band gaps obtained for R→Γ are 1.87 and 3.16 eV respectively.

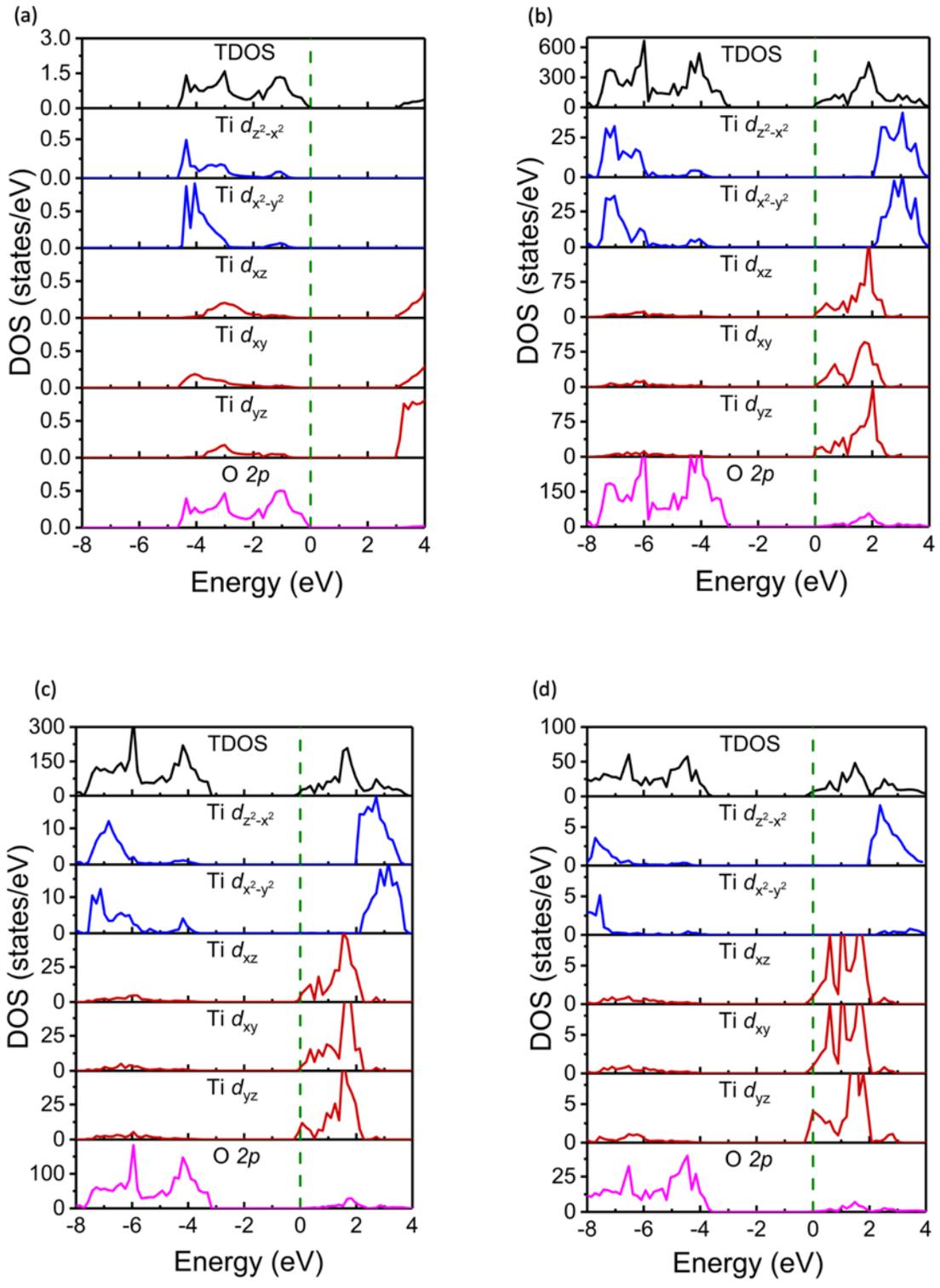

**Figure 3**: DOS and PDOS of a) Pure b) 1.6% c) 3.7 % and d) 12.5 % Ta doped STO denoting the shift of the Fermi level (shown by green dashed lines) deep inside the conduction band.

Here, we first focus on the change in the density of states (DOS) of STO due to doping. As shown in Figure 3(a), the valence bands of pure STO near the Fermi level (indicated by dashed green-line) is dominated by the O *2p*-Ti *3d* hybridized orbitals, while the bottom of the conduction bands is primarily governed by the empty Ti-*3d* states, with a very small contribution from the O-*2p* states. The Sr-*3d* orbitals, however, have been found not to contribute to total DOS near the Fermi level, in accordance with the nearly full ionic character of the Sr ions. In other words, the Sr atoms only participate in creating the perovskite structure but do not affect the electronic structure of STO around the Fermi level. Our DOS studies highlight that STO undergoes transition from insulating-to-semiconducting state under doping of Ta. As discussed above, the introduction of Ta into the crystal of STO contributes electrons to the system. These free electrons fill up the low-lying empty conduction bands of STO, which makes the doped systems into a *n*-type semiconductor. Because of doping, the Fermi level shifts into the conduction band and progressively embeds itself deeper with increasing doping levels, as shown in Fig. 3 (b), (c), and (d). As a result, the electrons now need to gain an extra energy $\Delta E$ (see Fig. 4(a)) for transition from the VBM to the empty conduction states, which is familiarly known as the Moss-Burstein shift [39,40]. Electrons present in the lower level of the conduction bands (i.e., below the Fermi level) can also jump to the empty states of the conduction band. These transitions are known as intra-band transitions and are responsible for an increase of the metallic conductivity of material. The increase of inter band-gap, i.e., the gap between the VBM and the Fermi level as well as $\Delta E$, as a function of doping are demonstrated in Figure 4(b). While it may appear that both inter- and intra-band gap increases linearly with rising doping concentrations, this is not the case in reality due to the strong influence of selection rules on electron transitions as well as electron-electron interactions that arises due a large number of free carriers in the system. This is clearly seen in Figure 4(b), that as the doping concentration reaches ~10%, $E_{opt}$ as well as $\Delta E$, begins to saturate at around a constant value of ~3.5eV and 0.2eV. This saturation of band gap value cannot be explained by band structure studies alone, and studying the optical band gap of the materials through frequency dependent dielectric function becomes essential to accurately understand the variation of band gap under doping. This aspect has been examined in detail in the next section through a combination of theoretical analysis and experimental measurements.

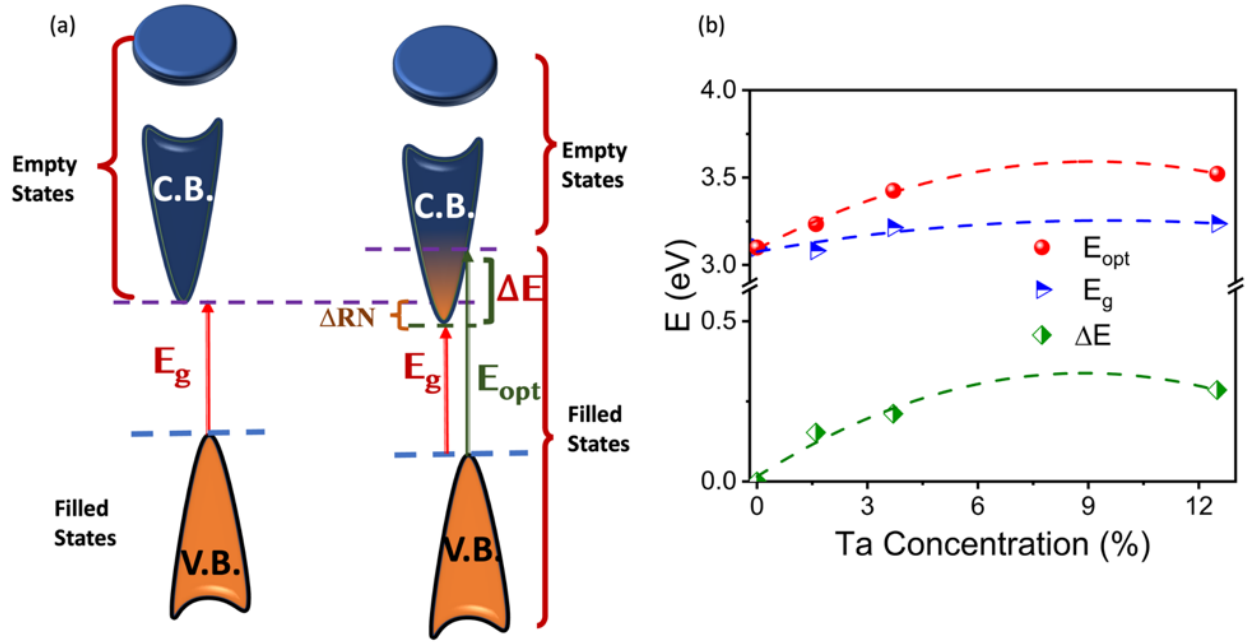

**Figure 4:** (a) Left side shows the band gap for pure semiconductors and the Moss-Burstein shift of doped semiconductors is seen on the right side. It clearly depicts the increase of band gap due to the introduction of electrons in the conduction bands. (b) Variation in the band gap as a function of the doping for the two different exchange correlation functional.

Moreover, our study of the partial density of states (*p*DOS) of STO reveals that the d-orbitals of Ti splits into triply degenerate $t_{2g}$ ($d_{xy}$, $d_{yz}$ and $d_{zx}$) and doubly degenerate $e_g$ ($d_{z^2-x^2}$ and $d_{x^2-y^2}$) orbitals, with latter orbitals situated higher in energy than former. The experimental confirmation of such splitting is evident from XAS [60–62] and PIXE [63] investigations conducted on STO thin films and single crystals. The crystal field splitting is also seen in the doped systems as well. However, in this case, the Fermi level intersects the low lying $t_{2g}$ states of Ti and Ta, as seen in Fig. 3 (b), (c), and (d). This suggests that the metallic conductivity of the doped system arises from the electrons filling in the $t_{2g}$ states of Ti and Ta. The *p*DOS for Ta shown in in Figure S4.

We now focus on investigating the modification in the nature of the band gap of STO resulting from the doping of Ta. It is known that the incorporation of supercell in DFT calculations introduces some spurious bands in the band dispersion due to the band-folding effect. In this case, it becomes necessary to unfold the DFT band structure to accurately determine the nature of band gap. In order to overcome this shortcoming, we utilize the BandUP [46,47] code to unfold the band structures of the examined doped systems. Given that the PBE functional provides an accurate depiction of band dispersion of STO, we rely on the same functional in this context as seen in Figure S5. The unfolded band structures of 2×2×2, 3×3×3, and 4×4×4 Ta doped supercells of STO are shown in Figure 5 (a), (b), and (c), respectively. There are several reports that suggest a change from indirect to direct band gap of STO through doping, strain, or vacancy effects [21,64]. However, in our study, we do not observe such a change due to doping of Ta. As can be seen in Figure 5, the introduction of dopants to the STO lattice introduces a few new bands in the band structures. However, since these new bands are far from the Fermi level, they have no effects on the effective mass of electrons near the CBM and hence to the mobility. This was clearly demonstrated in the case of Nb/ La doped STO, another n-type STO, where mobilities and effective mass of this doped system undergoes a minimal change for low to moderate doping levels of Nb [12,65]. Moreover, the ARPES study on Nb doped STO has also been found not to show a significant change in the band structure, as compared to the pristine system [29]. Thus, it is expected that the mobility and the nature of band dispersion of low Ta doped STO will remain consistent as the pristine system. In fact, such changes have been reported by Yaremchenko et al. [25], where they found no significant change in the mobility (~5 $cm^2V^{-1}s^{-1}$) and effective mass (~ 1.2 $m_e$) up to 20% Ta doped STO, in agreement to our previous report [26,27].

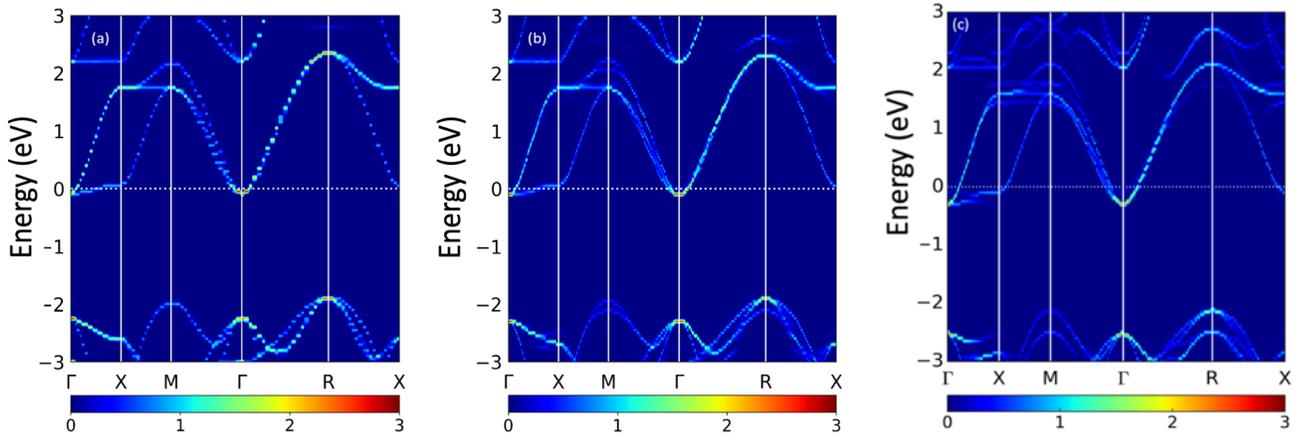

**Figure 5**: Unfolded band structure of different concentrations (a) 1.6 %, (b) 3.7%, (c) 12.5%. The color spectrum at the bottom of each represents the spectral weight density of electrons and are required to unfold the band structure effectively (See Medeiros et al. for theory [47]). The extra bands have very low spectral density, but the CBM and VBM have same spectral density as for pure STO.

### C. Optical characterization

Before commenting on the nature and evolution of band gaps obtained in the above section, we first determine the optical band gaps experimentally. For this, epitaxial thin films of $SrTi_{1-x}Ta_xO_3$ (x = 0, 0.02, and 0.05) are fabricated on LAO substrates with the help of PLD. The thicknesses of the thin films obtained through RBS is ~100 - 130nm respectively. During growth of the thin films, the partial pressure ($pO_2$) is kept constant at $5×10^{-5}$ mbar to remove any unwanted defects and vacancies that arises during the film growth process. This helps to maintain the stoichiometry of all the samples even with the doped samples. Our previous studies as well as other reports have shown that the samples do not exhibit a large compressive strain, which might be introduced due to the substrate (LAO, a ~ 3.79 Å), and are already relaxed at thicknesses above 80 nm. The thin films are thus expected to be in cubic-type phase, the same structure that we have used for the theoretical calculations.

The optical characteristics of the thin films are verified through the use of spectroscopic ellipsometer (SE). SE is a non-destructive technique that can provide information on the thickness, optical functions, surface roughness, interface features in much more detail than ordinary methods. The fundamental equation in ellipsometry is given by [66–69],

$$\rho = \frac{r_p}{r_s} = \tan\psi \cdot e^{i\Delta} \tag{2}$$

where $r_s$ and $r_p$ represent the complex Fresnel coefficients of perpendicular and parallel polarized light. $\rho$ is the complex reflectance ratio; $\psi$ describes the ratio change in amplitude between the two polarizations, and $\Delta$ represents the phase difference between them. Only the quantities $\psi$ and $\Delta$ are measured by ellipsometry. The complex interaction of electromagnetic waves with the sample is all contained within the two components. Measurements are taken at different angles to average out the data and to obtain consistent properties. Moreover, the extraction of correct optical constants requires some background knowledge of the samples regarding their electronic properties, reflectance, and absorbance etc. Then the data is fitted by a particular model (or a mix of several models) through non-linear regression algorithms and varying the fitting parameters such that the root mean square error (MSE) between the fitted and experimental data is as low as possible. After fitting the data properly, the optical functions obtained are n, k, $\varepsilon_1$, and $\varepsilon_2$. Here, n and k are the refractive index and extinction coefficient, respectively, whereas $\varepsilon_1$ and $\varepsilon_2$ are the real and imaginary parts of the dielectric function $\tilde{\varepsilon}$. They are all interconnected to each other by the following equations [67,68]:

$$\tilde{n} = n \pm ik \tag{3}$$

$$\tilde{\varepsilon} = \varepsilon_1 \pm i\varepsilon_2 \tag{4}$$

and,

$$\tilde{\varepsilon} = \tilde{n}^2 \tag{5}$$

Furthermore, the absorption coefficient, $\alpha$ can be calculated with the help of k, given by the equation $\alpha = 4\pi k/\lambda$ where $\lambda$ is the wavelength of the light.

The raw data collected through SE ($\psi$ and $\Delta$) can be modelled by various fitting models available. In this work, a modified form of Lorentz Oscillator, namely Tauc-Lorentz (TL) Oscillator is utilized to fit the data. The TL model incorporates [70,71] an extra term of TL band gap ($E_g$) apart from the four terms present in Lorentz oscillators, namely amplitude parameter (A), broadening parameter (C), peak transition energy ($E_{(n_0)}$) and energy independent contribution to $\varepsilon_1(E)$. Due to the inclusion of the band gap term, this model is shown to be implemented successfully for transparent and slightly absorbing semiconductors and provides a correct fit to the complex refractive index, which is necessary to extract all other important parameters required for complete studies of the system.

However, the introduction of free carriers due to doping renders the TL model insufficient to describe the optical properties of the system. Moreover, for metals and metallic systems, which have free charge carriers, generally the Drude model is successful in explaining the effects of free carrier effects. For relatively high doped systems, thus, a combination of Drude and TL oscillators are utilized to understand the complexity of the doped systems. This approach has been performed successfully in the past for pure and doped ZnO and is shown to be consistent with the experimental observations [70]. The dielectric function that combines the two oscillators Drude ($\varepsilon_D$) and TL ($\varepsilon_{TL}$) has the equation given by [70]:

$$\varepsilon(E) = \varepsilon_D + \varepsilon_{TL}(E) \tag{6}$$

where the contribution from the two terms is

$$\varepsilon_D = -\frac{A_D}{E^2 - i\Gamma_D E} = \left(-\frac{A_D}{E^2 + \Gamma_D^2}\right) - i\left(-\frac{A_D \Gamma_D}{E^3 + \Gamma_D^2 E}\right) \tag{7}$$

where $A_D$, $G_D$, and $\Gamma_D$ depicts amplitude, broadening and scattering time, respectively. The equation for TL oscillator is $\varepsilon_{TL} = \varepsilon_1 \pm i\varepsilon_2$, where

$$\varepsilon_2 = \frac{AE_{(n_0)}C(E_n - E_g)^2}{(E_n - E_{(n_0)})^2 + C^2 E_n^2} \frac{1}{E_n} \quad (E_n > E_g) \tag{8a}$$

$$\varepsilon_2 = 0 \; (E_n \leq E_g) \tag{8b}$$

and the real part of the dielectric expression $\varepsilon_1(E)$ can be derived from the Kramers-Kronig relation (KKR), given both $\varepsilon_1$ and $\varepsilon_2$ are KKR consistent.

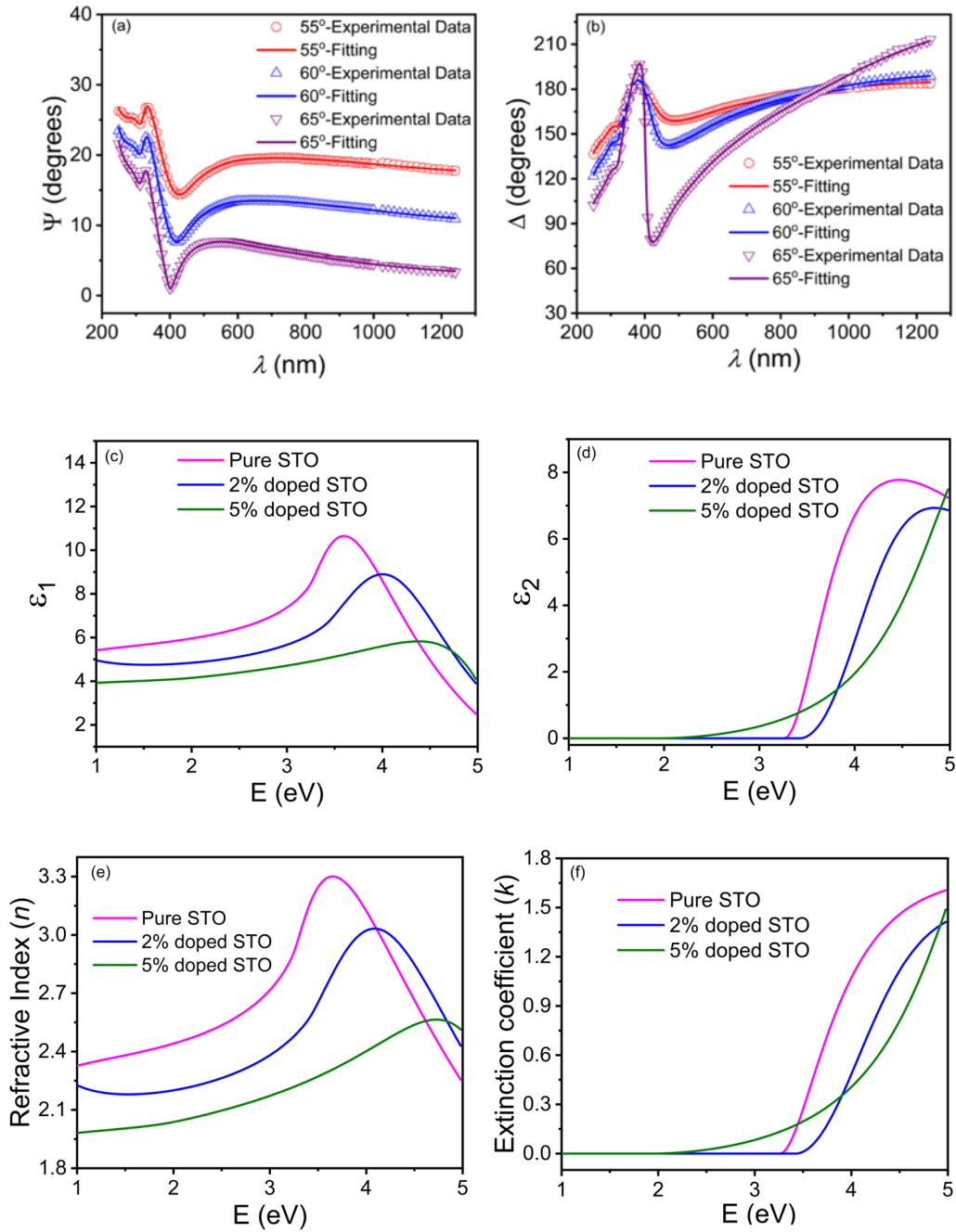

**Figure 6:** Representative ψ and Δ of the 2% Ta doped sample. The open structures represent raw data and the line represent the fitting. The angles are colour coded with the fitting line. (c) and (d) represents the real and imaginary part of the dielectric function. (e) and (f) represents the refractive index and extinction coefficient.

The spectroscopic ellipsometry fitting and extracted parameters obtained are given in Figure 6. Figure 6 (a) and (b) represents the $\psi$ and $\Delta$ of 2% doped STO while other samples are shown in supplementary Figure S6. The data is taken for the wavelength range of 248-1240 nm (or 1-5 eV) at three different angles of 55°, 60°, and 65°. The raw data is represented by the open circles and the fitting is shown by a line through the open circles in same color. A MSE less than 5% is obtained for each sample with the thickness deviation from RBS measurements not more than ± 5 nm. To qualitatively discuss the properties of the samples, roughness of the sample is also considered while fitting the data. The average roughness obtained for each is ~2 nm, compatible with the AFM data (not shown here. See Ref. [26]). The real and imaginary part of the dielectric function is represented in Figure 6(c) and (d). Correspondingly, the refractive index (n) and extinction coefficient (k) is shown in Figure 6 (e) and (f).

The $\varepsilon_1$ of pure sample that is obtained here matches very well with the previous reports [72]. As seen in Figure 6 (c) and (e), the peak position of $\varepsilon_1$ (and *n*) shifts towards a higher energy value as a function of doping. The peak position of the samples in these figures shows the region of anomalous behaviour, i.e., the region around which interband transitions take place. The shift towards higher energy side signifies that a higher energy of excitation is required for the interband transition. It is also seen that the absolute value of these terms ($\varepsilon_1$ and *n*) decreases upon doping. This is the characteristic of metallic systems, which have lower refractive index compared to semiconductor or insulating systems [70]. Both these studies prove that the electronic conductivity and the band gap increases as a function of doping, consistent with the B-M effect and the theoretical results obtained earlier. The imaginary part of the dielectric effect $\varepsilon_2$ and extinction coefficient (*k*) of the dielectric effect shows the shift of peaks towards higher energy side, signifying an increase in the band gap, consistent with the previous behaviour.

While the band structure studies shown above are necessary to understand the indirect nature of the Ta doped STO band transitions, they still do not provide complete information on the optical properties that we obtained from spectroscopic ellipsometry. To observe the complete optical transitions, such as the effect of intraband transitions, or free carrier absorption (FCA), and its effect on the optical transmission, frequency dependent optical properties are calculated for pure and doped STO systems using HSE06 xc as implemented in VASP. The absorption coefficient for pure and doped STO system obtained theoretically and experimentally are shown in Figure S7. As seen from the graph, the introduction of Ta on STO shifts the first absorption peak to the higher energies, meaning a widening of band gap. The effect is minute till 3.7 % followed by a large jump for 12.5% doping. However, near low energies (below 1eV) a surprising change is seen. The absorption peak intensity here increases a function of doping, and for 12.5% doping, the absorption peak spectrum increases appreciably (inset Figure S7(a)). This effect may arise due to the intraband transitions between the filled level and the empty levels of conduction band due to excess electrons in the system and leads to FCA [73]. This effect is widely known and seen in samples exhibiting very high metallicity such as $SrNbO_3$ [74], $SrMoO_3$, and $SrVO_3$ [75,76] etc. The FCA leads to an increase in the absorbance below the band edge [73], which results in the deterioration of the optical transmittance for heavily doped STO.

The unfolded band structure obtained earlier now helps to determine the value of n for calculating the band gap of the samples using the Tauc's plot, given by the equation:

$$\alpha h\nu = A\,(h\nu - E_g)^{1/n} \qquad (9)$$

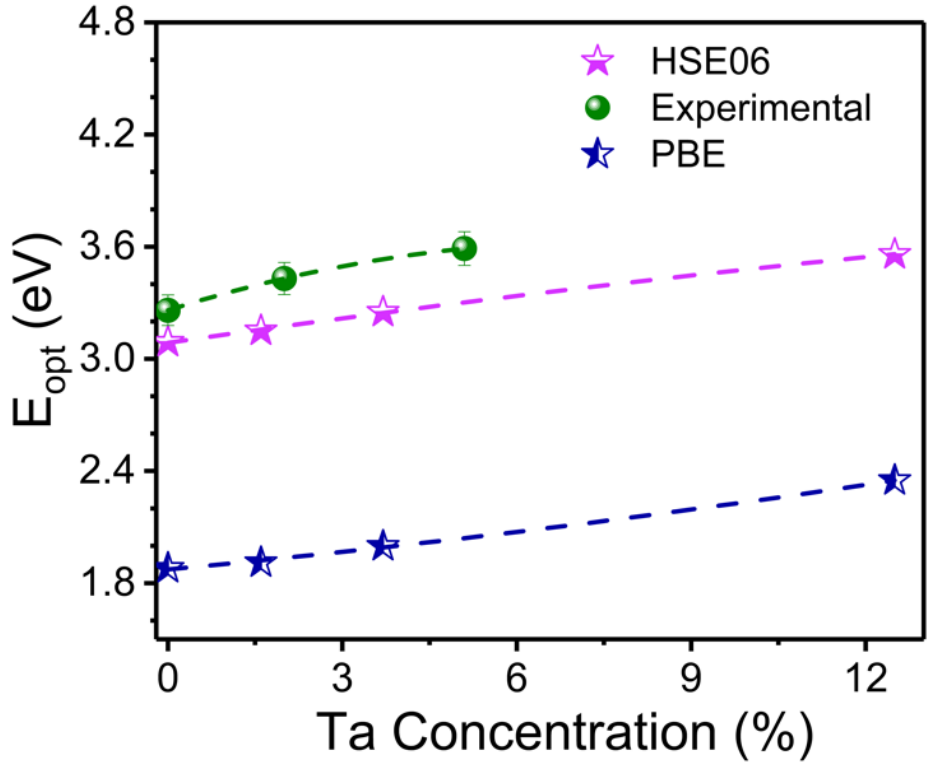

**Figure 7:** (a) Indirect band gap of Ta doped STO obtained by ellipsometry (b) The band gap variation of doped STO as function of carrier concentration obtained through all the studies.

The theoretical studies prove that the band structure remains indirect even up to high doping. As a result, the value of n is taken to be ½ for the calculation of the band gap. The trends of the band gaps obtained theoretically and experimentally show excellent agreement. For the sake of completion, we have also shown the variation of the band gaps from PBE XC, which agree with the trends observed above.

Upon doping of Ta in STO, the electronic conductivity of $SrTi_{1-x}Ta_xO_3$ increases due to the introduction of an extra electron in the system. The band gap variation is shown to have a non-linear feature. The band gap increase seems to saturate at a constant value at the high doping concentration, both theoretically as well as experimentally. As mentioned earlier, due to the alloy formation of $SrTi_{1-x}Ta_xO_3$ (x = 0 - 0.125), the band gap variation is not linear but follows Vegards law in a non-linear manner. Such studies on the influence of doping on A and B sites [77] of $BaTiO_3$-$BaZrO_3$ and $BaTiO_3$-$CaTiO_3$ systems have shown that first the band gap increases, followed by a miscibility region where it remains almost constant, followed by the region where characteristics of the doped alloy mixture dominates. This is due to two competing phenomena of Moss-Burstein shift and the band gap renormalization on the doped semiconductors [9]. The band gap narrowing effect is caused by electron-electron and electron-ion interactions, as well as the fusion of the impurity band and conduction band. The band gap changes can then be calculated by the modified form of Vegards law [77]. In the present case the band gap saturates at around 12.5 % theoretically and is expected to remain almost constant till the dominance of $SrTaO_3$ takes over and a structural transition occurs. Further doping has no large effect on the band gap of STO, and a very high doping induces a phase transition to $SrTaO_3$, which can occur in nature in its $Sr_2Ta_2O_7$ orthorhombic phase. The band gap of $Sr_2Ta_2O_7$ calculated through DFT (PBE XC) is 3.5eV but is predicted to be an insulator.

Since high doping on a single site cannot yield a major improvement in the optoelectronic properties, we suggest other routes to increase the performance of doped STO as a TCO. The optoelectronic effects of Ta doped STO can further be enhanced by co-doping of La at Sr site as well. Nitride [78,79] and Carbon [80] doping at the oxygen site in $SrTiO_3$ is shown to be of fantastic use in photocatalytic properties but decreases the band gap. However, the effects of doping due to two ions simultaneously can be complex due to the differences in ionic size, electron-defect interaction, electron-phonon interaction etc. to name a few and must be studied after the effects of co-doping are clear from first principles.

IV. **CONCLUSIONS**

The band gap evolution of the Ta doped STO in $SrTi_{1-x}Ta_xO_3$ is studied by combining the first principles method based on DFT and through the application of ellipsometry. We first change the concentration of Ta from 0-12.5% in the theoretical studies and study the effect of doping on the local site through bond length variation as well as its effect on lattice parameters. We then perform Bader charge analysis to understand the change of the oxidation state as a result of Ta doping. Next, we perform DOS and *p*DOS analysis on these systems to understand the evolution of STO from an insulator to a metallic system. The nature of the band gaps is verified by *band unfolding* method and has been found to have an indirect nature in all the cases, which is consistent with a few reported ARPES measurements on doped STO. To corroborate the findings of DFT, epitaxial thin films of pure and Ta doped STO are grown with Ta ranging from 0-5% through PLD. The films obtained are studied with the help of spectroscopic ellipsometry. A combination of TL and Drude oscillator methods are utilized to study the optical properties of Ta doped STO. The trend of the band gap variation obtained by SE matches well with the band gaps obtained earlier and has a non-linear behaviour. The optical properties of the system are then computed by frequency dependent dielectric calculations and are found to have the same behaviour obtained from ellipsometry. We thus conclude that when we dope STO with Ta, then apart from the Moss-Burstein shift, other effects such as band gap renormalization also occurs which counteracts the increase in the band gap of the system. As a result of this, the band gap value seems to saturate after a while and no band widening occurs further without inducing a significant structural transition. We also shed light on other routes to increase the optoelectronic performances of *n*-type STO for its industrial applications.

## ACKNOWLEDGEMENTS

We would like to thank the Shiv Nadar Institution of Eminence (Deemed to be University) for providing the research funding (Grant No. SNS/PHY/2013-20), as well as the DST-Science and Engineering Research Board (SERB) India for providing the funding (Grant No. SR/FST/PS-I/2017/6C). In order to perform theoretical calculations, the School of Natural Sciences at Shiv Nadar Shiv Nadar Institution of Eminence (Deemed to be University) makes use of its high-performance computing facility. In addition, we would like to express our gratitude to D. Das, and D. Chakraborty for their assistance and the insightful conversations they provided.

# REFERENCES

[1] X. Yu, T. J. Marks, and A. Facchetti, Metal oxides for optoelectronic applications, Nature Mater 15, 383 (2016).

[2] M. Azani, A. Hassanpour, and T. Torres, Benefits, Problems, and Solutions of Silver Nanowire Transparent Conductive Electrodes in Indium Tin Oxide (ITO)-Free Flexible Solar Cells, Adv. Energy Mater. 10, 2002536 (2020).

[3] Y. Ha, J. Byun, J. Lee, and S. Lee, Design Principles for the Enhanced Transparency Range of Correlated Transparent Conductors, Laser & Photonics Reviews 15, 2000444 (2021).

[4] A. Altuntepe, M. A. Olgar, S. Erkan, O. Hasret, A. E. Keçeci, G. Kökbudak, M. Tomakin, A. Seyhan, R. Turan, and R. Zan, Hybrid transparent conductive electrode structure for solar cell application, Renewable Energy 180, 178 (2021).

[5] C. Zhang, C. Ji, Y. Park, and L. J. Guo, Thin-Metal-Film-Based Transparent Conductors: Material Preparation, Optical Design, and Device Applications, Adv. Optical Mater. 9, 2001298 (2021).

[6] J. Shi, J. Zhang, L. Yang, M. Qu, D. Qi, and K. H. L. Zhang, Wide Bandgap Oxide Semiconductors: from Materials Physics to Optoelectronic Devices, Adv. Mater. 33, 2006230 (2021).

[7] C.-C. Wu, Ultra-high transparent sandwich structure with a silicon dioxide passivation layer prepared on a colorless polyimide substrate for a flexible capacitive touch screen panel, Solar Energy Materials and Solar Cells 207, 110350 (2020).

[8] B. Sarma, D. Barman, and B. K. Sarma, AZO (Al:ZnO) thin films with high figure of merit as stable indium free transparent conducting oxide, Applied Surface Science 479, 786 (2019).

[9] H. S. So, J.-W. Park, and D. H. Jung, Optical properties of amorphous and crystalline Sb-doped $SnO_2$ thin films studied with spectroscopic ellipsometry: Optical gap energy and effective mass, J. Appl. Phys. (2015).

[10] T. T. Werner, L. Ciacci, G. M. Mudd, B. K. Reck, and S. A. Northey, Looking Down Under for a Circular Economy of Indium, Environ. Sci. Technol. 52, 2055 (2018).

[11] C. Baranzelli et al., Methodology for Establishing the EU List of Critical Raw Materials: Guidelines (Publications Office of the European Union, Luxembourg, 2017).

[12] J. Ravichandran, W. Siemons, H. Heijmerikx, M. Huijben, A. Majumdar, and R. Ramesh, An Epitaxial Transparent Conducting Perovskite Oxide: Double-Doped $SrTiO_3$, Chem. Mater. 22, 3983 (2010).

[13] K. van Benthem, C. Elsässer, and R. H. French, Bulk electronic structure of $SrTiO_3$: Experiment and theory, Journal of Applied Physics 90, 6156 (2001).

[14] J. D. Baniecki, M. Ishii, H. Aso, K. Kurihara, and D. Ricinschi, Density functional theory and experimental study of the electronic structure and transport properties of La, V, Nb, and Ta doped $SrTiO_3$, Journal of Applied Physics 113, 013701 (2013).

[15] R. B. Comes, P. V. Sushko, S. M. Heald, R. J. Colby, M. E. Bowden, and S. A. Chambers, Band-Gap Reduction and Dopant Interaction in Epitaxial La,Cr Co-doped SrTiO$_3$ Thin Films, Chem. Mater. 26, 7073 (2014).

[16] S. Ahmed, T. Hasan, A. K. M. S. H. Faysal, S. S. Nishat, M. N. I. Khan, A. Kabir, and I. Ahmed, A DFT+U approach to doped SrTiO$_3$ for solar harvesting applications, Computational Materials Science 214, 111743 (2022).

[17] H. Bentour, M. El Yadari, A. El Kenz, and A. Benyoussef, DFT study of electronic and optical properties of (S–Mn) co-doped SrTiO$_3$ for enhanced photocatalytic hydrogen production, Solid State Communications 312, 113893 (2020).

[18] A. Raturi, P. Mittal, and S. Choudhary, Tuning the electronic and optical properties of SrTiO$_3$ for optoelectronic and photocatalytic applications by plasmonic-metal doping: a DFT-computation, Opt Quant Electron 54, 634 (2022).

[19] J. Wang, T. Fang, S. Yan, Z. Li, T. Yu, and Z. Zou, Highly efficient visible light photocatalytic activity of Cr–La codoped SrTiO$_3$ with surface alkalinization: An insight from DFT calculation, Computational Materials Science 79, 87 (2013).

[20] S. Winczewski, J. Dziedzic, T. Miruszewski, J. Rybicki, and M. Gazda, Properties of Oxygen Vacancy and Hydrogen Interstitial Defects in Strontium Titanate: DFT + U$^{d,p}$ Calculations, J. Phys. Chem. C 126, 18439 (2022).

[21] Y. S. Hou, S. Ardo, and R. Q. Wu, Hybrid density functional study of band gap engineering of SrTiO$_3$ photocatalyst via doping for water splitting, Phys. Rev. Materials 5, 065801 (2021).

[22] B. Modak and S. K. Ghosh, Enhancement of Visible Light Photocatalytic Activity of SrTiO$_3$ : A Hybrid Density Functional Study, J. Phys. Chem. C 119, 23503 (2015).

[23] Y. Liu, W. Zhou, and P. Wu, Electronic structure and optical properties of Ta-doped and (Ta, N)-codoped SrTiO$_3$ from hybrid functional calculations, Journal of Applied Physics 121, 075102 (2017).

[24] T. Moriga, R. Minakata, Y. Nomura, H. Ishikawa, K. Murai, and M. Mori, Stability and electrical conductivity of Nb- or Ta- doped SrTiO$_3$ perovskites for interconnectors in solid oxide fuel cells, J. Ceram. Soc. Japan 125, 223 (2017).

[25] A. A. Yaremchenko, S. Populoh, S. G. Patrício, J. Macías, P. Thiel, D. P. Fagg, A. Weidenkaff, J. R. Frade, and A. V. Kovalevsky, Boosting Thermoelectric Performance by Controlled Defect Chemistry Engineering in Ta-Substituted Strontium Titanate, Chem. Mater. 27, 4995 (2015).

[26] M. Arya, S. Kumar, D. Hasina, S. Ojha, A. Arora, V. K. Malik, A. Mitra, T. Som, and S. Dhar, Ta-doped SrTiO$_3$ epitaxial thin film: A promising perovskite for optoelectronics, Journal of Applied Physics 129, 145109 (2021).

[27] M. Arya, S. Kumar, D. Hasina, R. Sen, S. Ojha, V. Kumar, T. Som, and S. Dhar, Combining experimental and modelling approaches to understand the expansion of lattice parameter of epitaxial SrTi$_{1-x}$Ta$_x$O$_3$ (x = 0–0.1) films, Computational Materials Science 217, 111917 (2023).

[28] H.-C. Chen, C.-W. Huang, J. C. S. Wu, and S.-T. Lin, Theoretical Investigation of the Metal-Doped $SrTiO_3$ Photocatalysts for Water Splitting, J. Phys. Chem. C 116, 7897 (2012).

[29] M. Takizawa, K. Maekawa, H. Wadati, T. Yoshida, A. Fujimori, H. Kumigashira, and M. Oshima, Angle-resolved photoemission study of Nb-doped $SrTiO_3$, Phys. Rev. B 79, 113103 (2009).

[30] W. Ku, T. Berlijn, and C.-C. Lee, Unfolding First-Principles Band Structures, Phys. Rev. Lett. 104, 216401 (2010).

[31] A. B. Gordienko and A. V. Kosobutsky, Analysis of the electronic structure of crystals through band structure unfolding, Phys. Solid State 58, 462 (2016).

[32] M. Chen and M. Weinert, Layer k-projection and unfolding electronic bands at interfaces, Phys. Rev. B 98, 245421 (2018).

[33] V. Popescu and A. Zunger, Extracting E versus $\vec{k}$ effective band structure from supercell calculations on alloys and impurities, Phys. Rev. B 85, 085201 (2012).

[34] Y. Zhang, B. Fluegel, M. C. Hanna, A. Mascarenhas, L.-W. Wang, Y. J. Wang, and X. Wei, Impurity perturbation to the host band structure and recoil of the impurity state, Phys. Rev. B 68, 075210 (2003).

[35] Y. Ikeda, A. Carreras, A. Seko, A. Togo, and I. Tanaka, Mode decomposition based on crystallographic symmetry in the band-unfolding method, Phys. Rev. B 95, 024305 (2017).

[36] Z. Lebens-Higgins et al., Direct Observation of Electrostatically Driven Band Gap Renormalization in a Degenerate Perovskite Transparent Conducting Oxide, Phys. Rev. Lett. 116, 027602 (2016).

[37] A. Verma, A. P. Kajdos, T. A. Cain, S. Stemmer, and D. Jena, Intrinsic Mobility Limiting Mechanisms in Lanthanum-Doped Strontium Titanate, Phys. Rev. Lett. 112, 216601 (2014).

[38] C. A. Niedermeier, S. Rhode, K. Ide, H. Hiramatsu, H. Hosono, T. Kamiya, and M. A. Moram, Electron effective mass and mobility limits in degenerate perovskite stannate $BaSnO_3$, Phys. Rev. B 95, 161202 (2017).

[39] T. S. Moss, The Interpretation of the Properties of Indium Antimonide, Proc. Phys. Soc. B 67, 775 (1954).

[40] E. Burstein, Anomalous Optical Absorption Limit in InSb, Phys. Rev. 93, 632 (1954).

[41] R. B. Comes, S. Y. Smolin, T. C. Kaspar, R. Gao, B. A. Apgar, L. W. Martin, M. E. Bowden, J. B. Baxter, and S. A. Chambers, Visible light carrier generation in co-doped epitaxial titanate films, Appl. Phys. Lett. 106, 092901 (2015).

[42] G. Kresse and D. Joubert, From ultrasoft pseudopotentials to the projector augmented-wave method, Phys. Rev. B 59, 1758 (1999).

[43] J. P. Perdew, K. Burke, and M. Ernzerhof, Generalized Gradient Approximation Made Simple, Phys. Rev. Lett. 77, 3865 (1996).

[44] J. Heyd, G. E. Scuseria, and M. Ernzerhof, Hybrid functionals based on a screened Coulomb potential, The Journal of Chemical Physics 118, 8207 (2003).

[45] J. Heyd, G. E. Scuseria, and M. Ernzerhof, Erratum: "Hybrid functionals based on a screened Coulomb potential" [J. Chem. Phys. 118, 8207 (2003)], The Journal of Chemical Physics 124, 219906 (2006).

[46] P. V. C. Medeiros, S. Stafström, and J. Björk, Effects of extrinsic and intrinsic perturbations on the electronic structure of graphene: Retaining an effective primitive cell band structure by band unfolding, Phys. Rev. B 89, 041407 (2014).

[47] P. V. C. Medeiros, S. S. Tsirkin, S. Stafström, and J. Björk, Unfolding spinor wave functions and expectation values of general operators: Introducing the unfolding-density operator, Phys. Rev. B 91, 041116 (2015).

[48] T. Ma, R. Jacobs, J. Booske, and D. Morgan, Understanding the interplay of surface structure and work function in oxides: A case study on $SrTiO_3$, APL Materials 8, 071110 (2020).

[49] A. Jain et al., Commentary: The Materials Project: A materials genome approach to accelerating materials innovation, APL Materials 1, 011002 (2013).

[50] T. Hasan, A. Saha, M. N. I. Khan, R. Rashid, M. A. Basith, M. S. Bashar, and I. Ahmed, Structural, electrical, and magnetic properties of Ce and Fe doped $SrTiO_3$, AIP Advances 12, 095003 (2022).

[51] E. Zhou, J.-M. Raulot, H. Xu, H. Hao, Z. Shen, and H. Liu, Structural, electronic, and optical properties of rare-earth-doped SrTiO3 perovskite: A first-principles study, Physica B: Condensed Matter 643, 414160 (2022).

[52] S. A. Azevedo, J. A.S. Laranjeira, J. L.P. Ururi, E. Longo, and J. R. Sambrano, An accurate computational model to study the Ag-doping effect on $SrTiO_3$, Computational Materials Science 214, 111693 (2022).

[53] F. Maldonado, L. Maza, and A. Stashans, Electronic properties of Cr-, B-doped and codoped $SrTiO_3$, Journal of Physics and Chemistry of Solids 100, 1 (2017).

[54] A. M. Kamerbeek, T. Banerjee, and R. J. E. Hueting, Electrostatic analysis of n-doped $SrTiO_3$ metal-insulator-semiconductor systems, Journal of Applied Physics 118, 225704 (2015).

[55] M. T. Gray, T. D. Sanders, C. A. Jenkins, P. Shafer, E. Arenholz, and Y. Suzuki, Electronic and magnetic phenomena at the interface of $LaAlO_3$ and Ru doped $SrTiO_3$, Applied Physics Letters 107, 241603 (2015).

[56] Y. Wei, J. Wan, J. Wang, X. Zhang, R. Yu, N. Yang, and D. Wang, Hollow Multishelled Structured $SrTiO_3$ with La/Rh Co-Doping for Enhanced Photocatalytic Water Splitting under Visible Light, Small 17, 2005345 (2021).

[57] Y. Xu, Y. Liang, Q. He, R. Xu, D. Chen, X. Xu, and H. Hu, Review of doping $SrTiO_3$ for photocatalytic applications, Bull Mater Sci 46, 6 (2022).

[58] A. A. B. Baloch, S. M. Alqahtani, F. Mumtaz, A. H. Muqaibel, S. N. Rashkeev, and F. H. Alharbi, Extending Shannon's ionic radii database using machine learning, Phys. Rev. Materials 5, 043804 (2021).

[59] E. Ertekin, V. Srinivasan, J. Ravichandran, P. B. Rossen, W. Siemons, A. Majumdar, R. Ramesh, and J. C. Grossman, Interplay between intrinsic defects, doping, and free carrier concentration in $SrTiO_3$ thin films, Phys. Rev. B 85, 195460 (2012).

[60] G. Berner et al., $LaAlO_3$/$SrTiO_3$ oxide heterostructures studied by resonant inelastic x-ray scattering, Phys. Rev. B 82, 241405 (2010).

[61] M. Salluzzo et al., Orbital Reconstruction and the Two-Dimensional Electron Gas at the $LaAlO_3$/$SrTiO_3$ Interface, Phys. Rev. Lett. 102, 166804 (2009).

[62] J.-S. Lee, Y. W. Xie, H. K. Sato, C. Bell, Y. Hikita, H. Y. Hwang, and C.-C. Kao, Titanium $d_{xy}$ ferromagnetism at the LaAlO$_3$/SrTiO$_3$ interface, Nature Mater 12, 703 (2013).

[63] C.-C. Chiu et al., Presence of Delocalized Ti 3d Electrons in Ultrathin Single-Crystal SrTiO$_3$, Nano Lett. 22, 1580 (2022).

[64] L. Wang, W. Pan, W. X. Hu, and D. Y. Sun, Strain-induced indirect-to-direct bandgap transition in an np-type LaAlO$_3$ /SrTiO$_3$ (110) superlattice, Phys. Chem. Chem. Phys. 21, 7075 (2019).

[65] A. Kinaci, C. Sevik, and T. Çağın, Electronic transport properties of SrTiO 3 and its alloys: Sr$_{1-x}$La$_x$TiO$_3$ and SrTi$_{1-x}$M$_x$O$_3$ ( M = Nb , Ta ), Phys. Rev. B 82, 155114 (2010).

[66] N. K. Gupta, M. Kumar, A. K. Tiwari, S. S. Pal, H. Wanare, and S. A. Ramakrishna, Spectroscopic ellipsometry-based investigations into the scattering characteristics of topologically distinct photonic stopbands, Appl. Phys. Lett. 121, 261103 (2022).

[67] H. G. Tompkins and E. A. Irene, editors , Handbook of Ellipsometry (William Andrew Pub. ; Springer, Norwich, NY : Heidelberg, Germany, 2005).

[68] H. G. Tompkins and J. N. Hilfiker, Spectroscopic Ellipsometry: Practical Application to Thin Film Characterization (Momentum Press, New York, NY, 2016).

[69] X. Chen and E. Pickwell-MacPherson, An introduction to terahertz time-domain spectroscopic ellipsometry, APL Photonics 7, 071101 (2022).

[70] C. Bohórquez, H. Bakkali, J. J. Delgado, E. Blanco, M. Herrera, and M. Domínguez, Spectroscopic Ellipsometry Study on Tuning the Electrical and Optical Properties of Zr-Doped ZnO Thin Films Grown by Atomic Layer Deposition, ACS Appl. Electron. Mater. 4, 925 (2022).

[71] S. Chen, P. Kühne, V. Stanishev, S. Knight, R. Brooke, I. Petsagkourakis, X. Crispin, M. Schubert, V. Darakchieva, and M. P. Jonsson, On the anomalous optical conductivity dispersion of electrically conducting polymers: ultra-wide spectral range ellipsometry combined with a Drude–Lorentz model, J. Mater. Chem. C 7, 4350 (2019).

[72] K. Dorywalski, B. Andriyevsky, M. Piasecki, N. Lemee, A. Patryn, C. Cobet, and N. Esser, Ultraviolet vacuum ultraviolet optical functions for SrTiO$_3$ and NdGaO$_3$ crystals determined by spectroscopic ellipsometry, Journal of Applied Physics 114, 043513 (2013).

[73] X. Zhang, G. Shi, J. A. Leveillee, F. Giustino, and E. Kioupakis, Ab initio theory of free-carrier absorption in semiconductors, Phys. Rev. B 106, 205203 (2022).

[74] X. Xu, C. Randorn, P. Efstathiou, and J. T. S. Irvine, A red metallic oxide photocatalyst, Nature Mater 11, 595 (2012).

[75] M. Mirjolet, M. Kataja, T. K. Hakala, P. Komissinskiy, L. Alff, G. Herranz, and J. Fontcuberta, Optical Plasmon Excitation in Transparent Conducting SrNbO$_3$ and SrVO$_3$ Thin Films, Adv. Optical Mater. 2100520 (2021).

[76] M. P. Wells et al., Tunable, Low Optical Loss Strontium Molybdate Thin Films for Plasmonic Applications, Advanced Optical Materials 5, 1700622 (2017).

[77] S. Lee, R. D. Levi, W. Qu, S. C. Lee, and C. A. Randall, Band-gap nonlinearity in perovskite structured solid solutions, Journal of Applied Physics 107, 023523 (2010).

[78] A. Fuertes, Nitride tuning of transition metal perovskites, APL Materials 8, 020903 (2020).

[79] J. J. Brown, Z. Ke, T. Ma, and A. J. Page, Defect engineering for photocatalysis: from ternary to perovskite oxynitrides, ChemNanoMat 6, 708 (2020).

[80] M. V. Makarova, A. Prokhorov, A. Stupakov, J. Kopeček, J. Drahokoupil, V. Trepakov, and A. Dejneka, Synthesis and Magnetic Properties of Carbon Doped and Reduced $SrTiO_3$ Nanoparticles, Crystals 12, 9 (2022).

# Supplementary Information

# Unveiling the structural, chemical state, and optical band-gap evolution of Ta-doped epitaxial SrTiO$_3$ thin films using first-principles calculations and spectroscopic ellipsometry

**Contents:**

**Figure S1:** Supercell configuration used for the calculations in the main manuscript

**Figure S2:** DFT calculated lattice parameters and Vegards law for SrTi$_{1-x}$Ta$_x$O$_3$ solid solution

**Figure S3:** Variation of bond lengths as a result of doping

**Figure S4:** Density of states for Ta ion in STO lattice

**Figure S5:** DFT calculated and unfolded effective band structure for pure STO in 2×2×2 supercell configuration

**Figure S6:** Ellipsometric raw data and fitting for pure and 5% Ta doped STO

**Figure S7:** DFT calculated and experimentally obtained absorption spectra for pure and doped STO

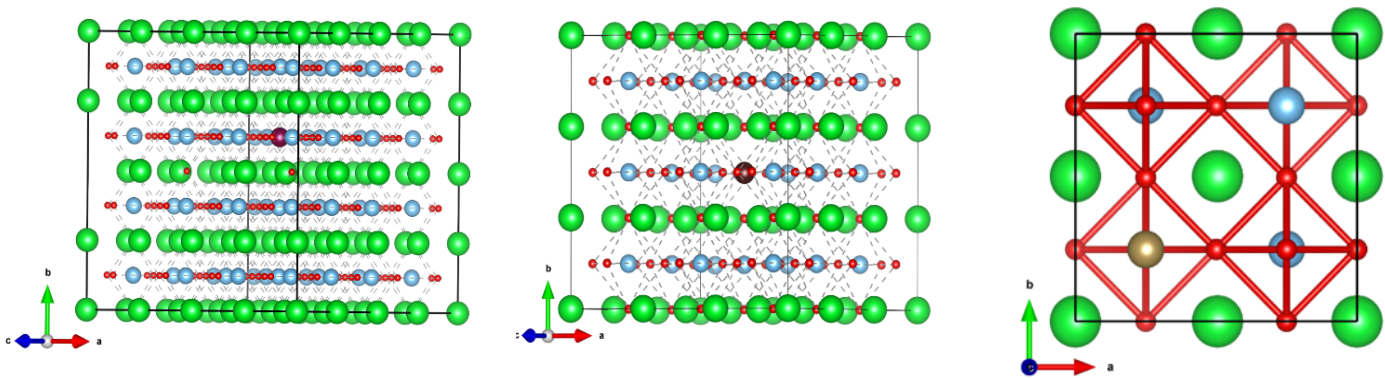

**Figure S1**: various supercell configurations of a) 4×4×4 with 1.6 % doping, b) 3×3×3 with 3.7% doping and c) 2×2×2 with 12.5 % doping.

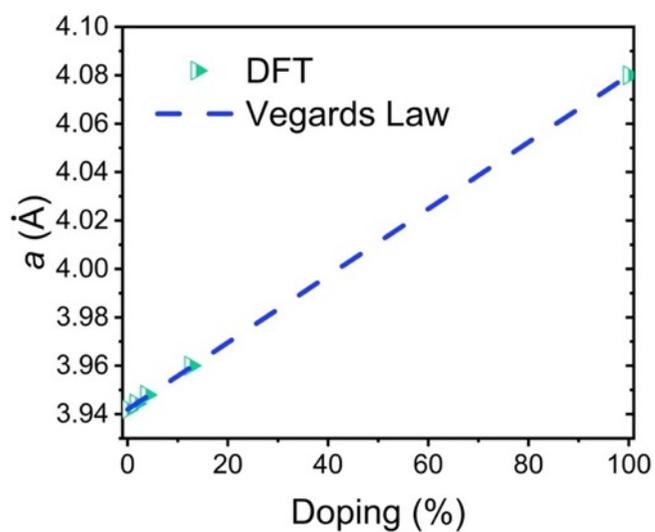

**Figure S2**: DFT calculated and Vegards law calculated lattice parameter. The Vegards law is given by $a_{Ta,STO} = (1-x) a_{STO} + x a_{STaO}$, where x represents doping and STaO represents $SrTaO_3$. The lattice parameter for STaO obtained through DFT is 4.08Å.

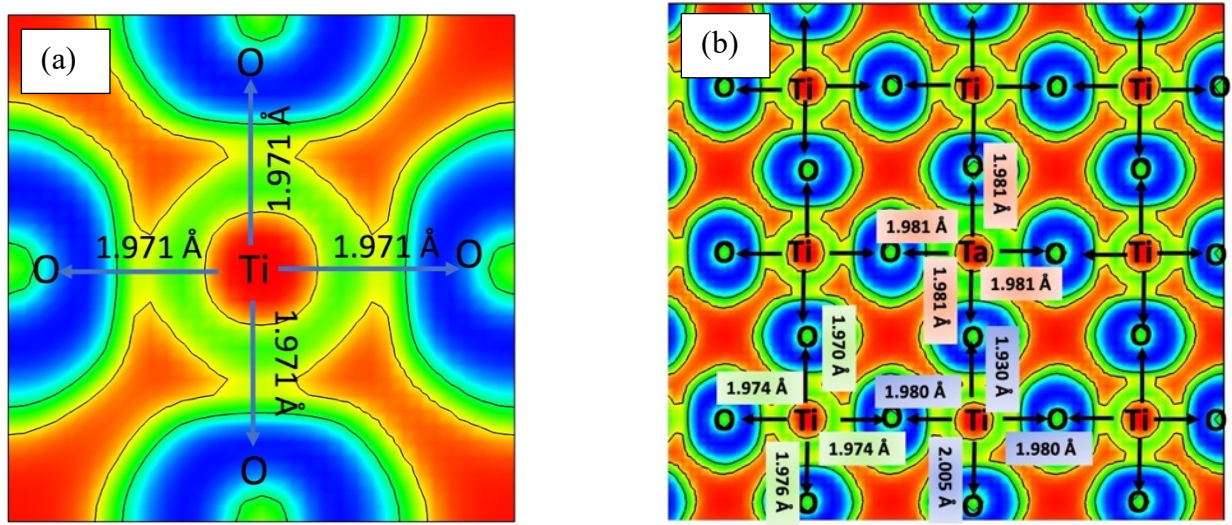

**Figure S3:** The change in the bond lengths of Ti-O for (a) pure and (b) 1.6% doped STO. In (b) the first and second neighbour is shown by blue and green text boxes, respectively. It is clearly seen from (b) that when the Ta is doped at the Ti site, the nearest Ti-O bond is shrinked (1.930Å) and stretched (2.005 Å) along the direction towards and away from the defect center and is also unequal along axial and equatorial directions. This distortion becomes less as we move away from the defect. Similar behaviour is observed in other doped cases as well.

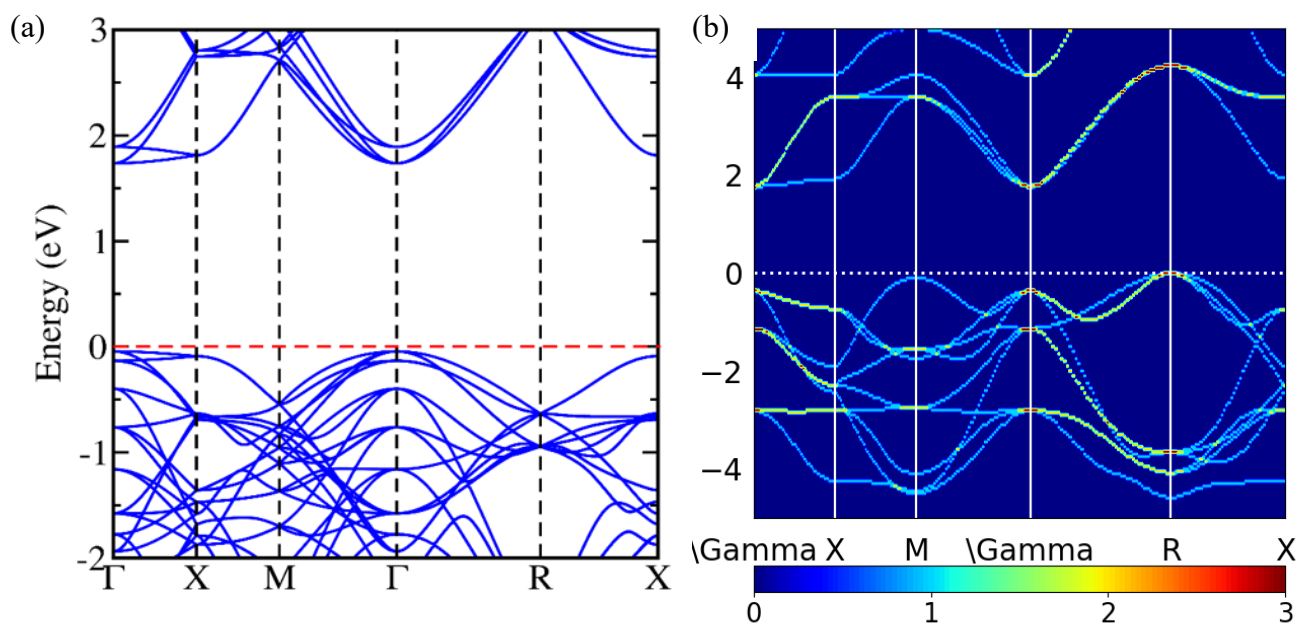

**Figure S4:** (a) DFT band and (b) unfolded bandstructure of 2×2×2 supercell of $SrTiO_3$. While DFT gives the direct nature, (b) upon unfolding the band structure retains the structure of pure $SrTiO_3$ unit cell.

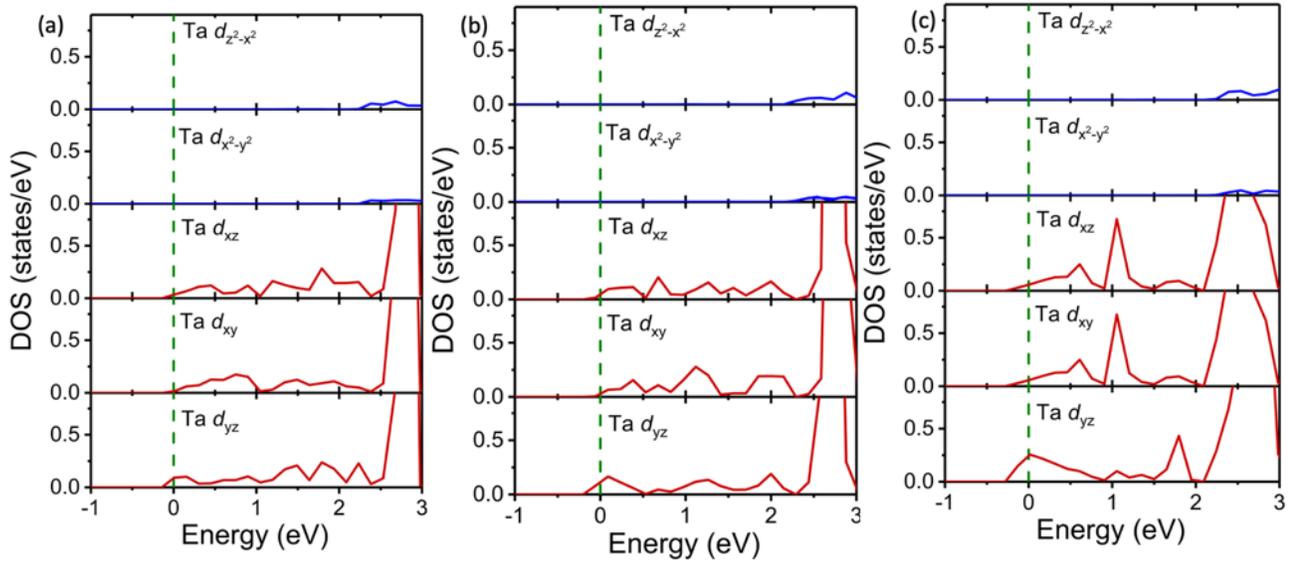

**Figure S5:** The splitting of Ta into its $e_g$ and $t_{2g}$ states for (a) 1.6% (b) 3.7% and (c) 12.5% doped cases. We have only shown the Ta contribution around the fermi level for simplicity.

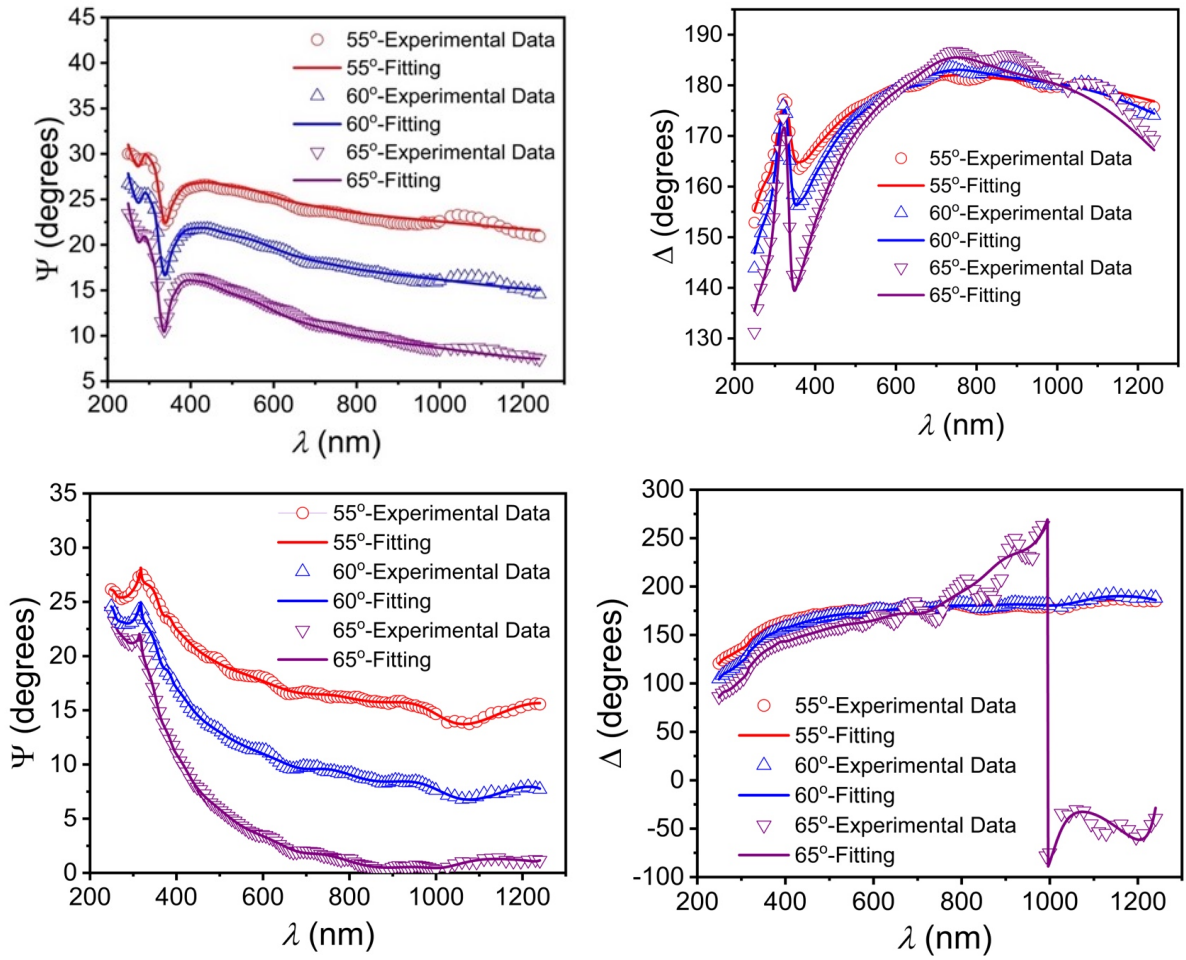

**Figure S6:** Ellipsometric raw data Psi for (a) pure and (c) 5% doped STO and delta for (b) pure and (d) 5% STO along with the fitted lines.

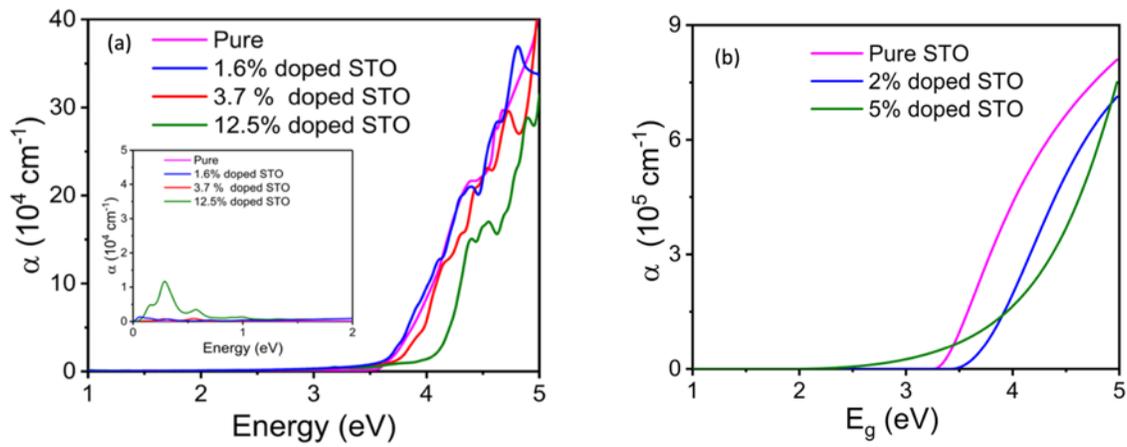

**Figure S7:** Absorption coefficient obtained from (a) theoretical calculations, and (b) spectroscopic ellipsometry. Inset of (a) shows the absorption coefficient in the low energy range.